\documentclass{kluwer}

\newdisplay{guess}{Conjecture}
\usepackage{url}

\newfont{\mm}{msbm10}

\begin{document} 
\begin{article}
\begin{opening}

\title{Classifying extrema using intervals}
\author{Marek W.~\surname{Gutowski}} 
\runningauthor{Marek W. Gutowski}
\runningtitle{Classifying extrema}
\institute{Institute of Physics, Polish Academy of Sciences, 
Al.~Lotnik\'ow 32/46,\\ 02--668 Warszawa,
Poland\email{Marek.Gutowski@ifpan.edu.pl}}

\begin{abstract}
    We  present  a~straightforward  and  verified  method  of   deciding 
 whether the point $x^{\star}\in\hbox{\mm R}^{n}$,\ $n\geqslant 1$, such 
 that $\nabla f(x^{\star})=0$,\ is the  local  minimizer,  maximizer  or 
 just a~saddle point of a~real-valued function $f$.  The  method  scales 
 linearly with dimensionality of the problem and  never  produces  false 
 results.
\end{abstract}


\classification{ACM codes}{F.2.2, 
G.1.0, 
G.1.2, 
J.2} 

\end{opening} 

\section{Introduction}

    This work is motivated by the practical problem, encountered  during 
 our studies in physics of magnetic materials.\  Namely,  we  wanted  to 
 investigate properties of simple magnetic systems,  consisting  of  few 
 entities, called spins, and treated as classical (i.e. not quantum)  2- 
 or 3-dimensional vectors of unit length.\ The positions  of  spins  are 
 fixed in space (in crystal lattice, for example),  but  the  spins  are 
 free to rotate -- accordingly to the interactions between them  and  to 
 the strength and orientation of  the  external  magnetic  field.\  Each 
 geometrical configuration of the system is  characterized  by  a~single 
 number called the free  energy.\  It  is  the  so  called  free  energy 
 landscape what we are interested in: the  positions  of  (stable)  free 
 energy minima, the valleys between them and so on.\ The free energy  is 
 a~smooth function, defined  on  the  open  domain  spanned  by  angular 
 variables describing the orientations of all the  spins  involved.  The 
 interactions between spins, as well as the numbers  characterizing  the 
 external field (if any), are fixed parameters.

    Very similar problem  is  encountered  in  computational  chemistry, 
 where the so called reaction pathways need to be traced.

\section{Standard approach and its deficiencies}
    The exploration of the free energy  landscape  usually  begins  with 
 solving the system of simultaneous equations:
\begin{equation}
    \frac{\partial f\left(x_{1}, x_{2},  \ldots,  x_{n}\right)}{\partial 
 x_{j}} = 0,\ \quad\ j=1, 2, \ldots, n,
\label{first}
\end{equation}
where $n$ is the number of unknowns (variables).

    From now on  we  will  assume  that  the  system  (\ref{first})  has 
 finitely  many   solutions   $x_{1}^{\star},   x_{2}^{\star},   \ldots, 
 x_{p}^{\star}$, with $p<\infty$. We will not  discuss  the  potentially 
 possible degenerate case, $p=\infty$  (countable  or  not)  for  purely 
 physical reasons: each real system always settles in a state with  well 
 defined magnetization, at least after sufficiently long time.

    Once    the     set     $X^{\star}=\left\{x_{k}^{\star}:\     \nabla 
 f(x_{k}^{\star})=0,\ k=1,2, \ldots, p\right\}$\ is known, we can  start 
 the classification procedure.\ It  should  tell  us  which  members  of 
 $X^{\star}$ are minimizers,\ maximizers or  correspond  to  the  saddle 
 points of $f$.\ The usual approach is to investigate the properties  of 
 Hessian of the function $f$,\ calculated for each  $x_{k}^{\star}$\  in 
 turn.\ Positive definiteness of the  matrix  $H_{ij}=\frac{\partial^{2} 
 f\left(x_{1},  x_{2},  \ldots,  x_{n}\right)}{\partial  x_{i}  \partial 
 x_{j}}\left|_{x=x_{k}^{\star}}\right.$\  is  the~sufficient  (but  {\em 
 not\/}  necessary)  condition  for  $f$  to  have  a~local  minimum  at 
 $x=x_{k}^{\star}$.\ Checking whether $H$ is positively defined is  easy 
 in  dimension  $n=2$  and  is   routinely   presented   in   analytical 
 calculations performed `by hand', as can be seen in many  textbooks  on 
 magnetism.\ For $n=2$  the  process  reduces  to  finding  whether  the 
 expressions
\begin{equation}
    \frac{\partial^{2}             f\left(x_{k}^{\star}\right)}{\partial 
 x_{1}^{2}},\  \frac{\partial^{2}  f\left(x_{k}^{\star}\right)}{\partial 
 x_{2}^{2}}      \quad\hbox{\rm      and}\quad\       \frac{\partial^{2} 
 f\left(x_{k}^{\star}\right)}{\partial x_{1}^{2}}\cdot\frac{\partial^{2} 
 f\left(x_{k}^{\star}\right)}{\partial           x_{2}^{2}}            - 
 \left(\frac{\partial^{2}  f\left(x_{k}^{\star}\right)}{\partial   x_{1} 
 \partial x_{2}}\right)^{2}
\label{conditions}
\end{equation}
are all positive (all negative when searching for maximum).

    When the dimensionality  of  the  problem  gets  higher,  the  above 
 approach becomes more and more tedious,  requiring  the  evaluation  of 
 many expressions of increasing complexity --  determinants  of  various 
 minors of  the  matrix  $H$  \cite{Korn}.\  In  automated  computations 
 another approach may appear more efficient,\  namely  finding  all  the 
 eigenvalues  of  the   matrix   $H\left(x_{k}^{\star}\right)$\/.\   The 
 positiveness (negativeness) of {\em  all\/}  its  eigenvalues  is  also 
 a~sufficient  condition   for   $H\left(x_{k}^{\star}\right)$   to   be 
 positively (negatively) defined and, consequently, for  $x_{k}^{\star}$ 
 to be a~local minimizer (maximizer) of~$f$. Needless to say  that  both 
 approaches  are,  except  for   nearly   trivial   cases,   practically 
 unsuitable for hand calculations and we have to rely  on  computers  to 
 perform this task.

    However, both those approaches suffer from two serious problems. The 
 first one is inherent to automatic computations, performed with limited 
 accuracy.\ Every investigated point $x_{k}^{\star}$,\  $k=1,\ldots,p$,\ 
 is  already  known  only   approximately   and   so   is   the   matrix 
 $H\left(x_{k}^{\star}\right)$.\  Rounding  errors  accumulating  during 
 either  procedure  can  only  worsen  this  situation  leading  to  the 
 unreliable or even false results.

    The second possible  deficiency  has  nothing  to  do  with  limited 
 accuracy and is related rather to the properties of the  function  $f$. 
 Consider     for     example     $f\left(x_{1},     x_{2}\right)      = 
 x_{1}^{2}+x_{2}^{4}$     having     exactly     one     minimum      at 
 $x^{\star}=\left(0,0\right)$.\   One    can    easily    check,    that 
 $H\left(x^{\star}\right)$ is a singular $2\times 2$  matrix,  with  the 
 only  non-vanishing  element  $\partial^{2}  f/\partial  x_{1}^{2}=2$.\ 
 Since  the  matrix  is  diagonal  then  we  have  immediately  its  all 
 eigenvalues: $\lambda_{1}=2$,\ $\lambda_{2}=0$\ --  not  all  positive. 
 No conclusion concerning  $x^{\star}=(0,0)$  is  thus  possible  during 
 exact calculations.\ It is  interesting,  however,  that  in  automated 
 calculations    we     may     arrive     at     slightly     perturbed 
 $\tilde{x}^{\star}=(0,\delta)$,\  with  $\delta\ne   0$,\   as   a~sole 
 candidate for a local minimizer.\ Now the Hessian  is  diagonal  again, 
 with  $H_{11}=2$  and  $H_{22}=12x_{2}^{2}=12\delta^{2}$,  leading   to 
 different conclusions.\ Depending on the particular value of  $\delta$, 
 $H_{22}$ either remains equal to zero within the machine accuracy, like 
 before, or is positive. For example,  working  with  accuracy  of  $10$ 
 decimal  digits  we  may  have:  $\delta=10^{-4}$  and  $H_{22}=2\times 
 10^{-8} > 0$, while the relevant component  of  gradient  is  $\partial 
 f/\partial x_{2} =4x_{2}^{3}= 4\times 10^{-12}$  --  the  number  which 
 will be rounded down to exactly  zero  by  our  computer.\  Looking  at 
 those    two    numbers    one    is    tempted    to    think     that 
 $\tilde{x}^{\star}=(0,\delta)$  is  a~true  minimizer  for  $f$.  Maybe 
 $x^{\star}=(0,0)$  is  another  one,  for   some   reason   missed   by 
 gradient-calculating routine.

\section{The interval solution}
    Here we present a simple and elegant solution to our problem,  based 
 on properties of interval calculus. Let us  recall  the  definition  of 
 a~local minimum of a real function $f$ of $n$ variables:

\newproof{defin}{Definition}
                {\par\bigskip }
\begin{defin}
    We say that $f\left(x_{1}, x_{2}, \ldots, x_{n}\right)$ has a  local 
 minimum at $x^{\star}$ when
$$
    \exists_{\varepsilon>0}\   \forall_{x\in    {\mathcal    D}(f)}\quad 
 \left|\!\left|x-x^{\star}\right|\!\right|<\varepsilon       \Rightarrow 
 f\left(x^{\star}\right) < f\left(x\right),
$$
    where $\left|\!\left|\cdot\right|\!\right|$ is any norm  defined  in 
 $\hbox{\mm R}^{n}$,\ and ${\mathcal  D}(f)\subset\hbox{\mm  R}^{n}$  is 
 the domain of the function $f$.
\end{defin}

    In simple words: moving away from the point $x^{\star}$, but  within 
 the limited range $\varepsilon$, always leads to the  increase  of  the 
 function value compared to $f\left(x^{\star}\right)$.

\subsection{Wrong, naive test for minimum and why it fails}

    Let us try to make direct use of  the  definition  above  and  let's 
 evaluate    the    function    $f$    at    the    following    points: 
 $\left(x_{1}^{\star},              x_{2}^{\star},               \ldots, 
 x_{k}^{\star}-\varepsilon,       \ldots,        x_{n}^{\star}\right)$,\ 
 $\left(x_{1}^{\star},              x_{2}^{\star},               \ldots, 
 x_{k}^{\star}+\varepsilon,  \ldots,  x_{n}^{\star}\right)$,\  for  some 
 fixed (presumably small) value of  $\varepsilon$  and  $k=1,2,  \ldots, 
 n$,\ in hope of reaching the conclusion  concerning  the  character  of 
 $x^{\star}$ -- whether it  is  a~local  minimum,  maximum  or  a~saddle 
 point  of  $f$.\  It  is  well  known  that  this  procedure  may  only 
 accidentally produce the correct answer. The  main  reason  is  that  it 
 does {\em not\/} sample every possible direction  around  $x^{\star}$.\ 
 Last but not least --  $x^{\star}$  may,  and  usually  will,  slightly 
 differ from the true location of minimum.

\subsection{Interval test}
    The naive test produces incorrect results but, fortunately, we  know 
 why.\ Nevertheless its simplicity is  so  tempting  that  the  idea  of 
 improving it makes sense.\ All we have to  change  is  the  ability  to 
 test the behavior of the given function  in  every  possible  direction 
 with respect to the suspected point. To achieve this goal  we  need  to 
 construct a closed surface around  $x^{\star}$  and  simply  check  the 
 range of $f$ on this surface.

    Here are the necessary steps of the interval-oriented  algorithm  to 
 determine the character of each  point  $x_{k}^{\star}\in  X^{\star}$,\ 
 $k=1,2,   \ldots,p$,\   i.e.   satisfying    the    equation    $\nabla 
 f\left(x^{\star}\right)=0$:
\begin{enumerate}
\item
    initialization: set $k=1$,
\item
    fix the attention at point $x_{k}^{\star}\in  X^{\star}$.  Calculate 
 the reference value $V_{k}= f\left(x_{k}^{\star}\right)$,
\item
    determine  the  distances  between  $x_{k}^{\star}$  and  all  other 
 members of the set $X^{\star}$ and discover the shortest one, $D_{k}$,
\item
    set $\varepsilon=D_{k}/2$,
\item
    generate  $2n$  interval  boxes  around  $x_{k}^{\star}$  with   the 
 following properties:
\begin{itemize}[$\bullet$]
\item
    the center of each box is an  image  of  the  center  (midpoint)  of 
 $x_{k}^{\star}$ shifted by $+\varepsilon$ or $-\varepsilon$  along  the 
 consecutive coordinate axes,
\item
    the size of each box in each  direction  is  $2\varepsilon$,  except 
 for the shift direction in which the width of box is equal to zero.
\end{itemize}
\item
    evaluate $f$ over each newly created  box  obtaining  the  intervals 
 $F_{1}^{+}$,    $F_{1}^{-}$,    $F_{2}^{+}$,    \ldots,    $F_{n}^{+}$, 
 $F_{n}^{-}$,
\item
count the events:
\begin{itemize}[$\bullet$]
\item
    $N_{0}$:\ the intervals $V_{k}$ and  $F_{j}^{\spadesuit}$  ($j=1, 
 2, \ldots, n$,\ \ $\spadesuit\in\{+,-\}$) intersect,
\item
    $N_{>}$:\ $V_{k}>F_{j}^{\spadesuit}$\ (every real number taken  from 
 $V_{k}$ is greater than any number form $F_{j}^{\spadesuit}$), and
\item
$N_{<}$:\ $V_{k}<F_{j}^{\spadesuit}$ is true.
\end{itemize}
    The classification of $x_{k}^{\star}$ is following:
\begin{itemize}[$\bullet$]
\item
    $N_{<}=2n\ \Rightarrow\ f$ has a local minimum at $x_{k}^{\star}$,
\item
    $N_{>}=2n\ \Rightarrow\ f$ has a local maximum at $x_{k}^{\star}$,
\item
    $N_{>}\cdot N_{<}\ne 0\ \Rightarrow$\ there is  a  saddle  point  at 
 $x_{k}^{\star}$\ \ (inflection point if $n=1$),
\item
    otherwise the case is undecided.
\end{itemize}
\item
    set $k\leftarrow k+1$.\ If  $k\le  p$  then  repeat  the  procedure, 
 starting from step $2$ else finish.
\end{enumerate}

\section{Discussion and final remarks}

    The sketch of the algorithm makes no  clear  statement  whether  the 
 elements of the set $X^{\star}$ belong to $\hbox{\mm R}^{n}$ or  rather 
 to $\hbox{\mm IR}^{n}$ -- the set  of  all  $n$-dimensional  intervals. 
 For the idea itself, as presented here, it is not  an  issue  and  both 
 interpretations are almost equally  good.  This  is  because  we  don't 
 discuss the ways to  obtain  the  set  $X^{\star}$.  What  we  require, 
 however, is that $X^{\star}$ contains {\em all}  the  solutions  of  an 
 equation $\nabla f=0$ within the domain of interest. This is because we 
 have to be able to precisely separate every one of such  solution  from 
 every other member of $X^{\star}$.\ In machine calculations the  really 
 important thing is the knowledge of guaranteed bounds for  each  member 
 of $X^{\star}$ and the certainty that those  solutions  are  separable, 
 even after their uncertainties are taken into account.

    The proposed routine avoids the most important trap  of  the  naive, 
 incorrect approach. It effectively samples {\em all\/}  the  directions 
 around the suspected point $x^{\star}$ and therefore is in full  accord 
 with the definition of a~local minimum.\  This  is  because  the  trial 
 boxes constructed by the algorithm  make  a  complete  and  `air-tight' 
 surrounding of the suspected point $x^{\star}$.\ In other  words  every 
 straight line that passes through  $x^{\star}$  must  also  necessarily 
 intersect two surrounding boxes.\  The  continuity  of  $f$,  which  is 
 differentiable and therefore continuous, assures that our algorithm  is 
 correct.\ It is the remarkable property of the  interval  calculations: 
 the ability of executing infinite and uncountable number of  operations 
 in a single step.\ This feature makes possible  to  convert  the  naive 
 and essentially wrong algorithm into a powerful and reliable tool.

    It may come as a surprise that our $\varepsilon$  is  rather  large, 
 contrary to the regular use of this symbol, mostly  thought  as  `being 
 sufficiently small'  or  `no  matter  how  small'.  We  prefer  to  use 
 $\varepsilon$ this big for a~good reason: too small value is  dangerous 
 and vulnerable to the other trap, namely that $x^{\star}$  is  inexact. 
 On the other hand the bigger $\varepsilon$ is  the  wider  can  be  the 
 intervals $F_{j}^{\spadesuit}$ and therefore we may obtain  `undecided' 
 result too often. Our  prescription  sets  the  safe  upper  limit  for 
 $\varepsilon$ rather than treats it as the one and only correct  value. 
 If $\varepsilon$ had higher value then our surface could  contain  more 
 than one element of $X^{\star}$.

    In practice the set  $X^{\star}$  will  be  determined  by  interval 
 methods (because only those methods  guarantee  that  {\em  all\/}  the 
 candidates for extrema can be found within the domain of interest)  and 
 therefore each its member will have the form of a~small  box.\  Setting 
 $\varepsilon$ equal to the width of such box, or only slightly  higher, 
 is the first  thing  coming  to  the  mind.\  One  should  not  forget, 
 however, that the interval calculus usually  overestimates  the  ranges 
 of the functions. For this reason the range of $f$ calculated  for  the 
 single  face  of  such  a~small  box  is  likely  to   have   non-empty 
 intersection with range of $f\left(x_{k}^{\star}\right)$.\ It  is  even 
 certain, if the true minimizer happens to be located  at  the  face  of 
 $x_{k}^{\star}$  currently  investigated  rather  than  laying  at  its 
 midpoint.

    Making the surrounding boxes `thin' in one direction is  a~trick  to 
 circumvent the notorious overestimates of interval enclosures.\ But  it 
 is not perfect and larger values of  $\varepsilon$,  somewhere  between 
 $D_{k}/2$ and half of the width of  $x_{k}^{\star}$  should  be  used.\ 
 'Undecided'   members   of   $X^{\star}$   may    be    retried    with 
 $\varepsilon^{\prime}=\left(\varepsilon           +\frac{1}{2}\hbox{\rm 
 width}\left(x_{k}^{\star}\right)\right)/2$.\ Allowing $\varepsilon$  to 
 be smaller that the halved  width  of  $x^{\star}$  is  dangerous:  the 
 uncertainty of $x^{\star}$ will almost surely  produce  false  results, 
 if ever.\ The other built-in  feature  of  the  algorithm  is  implicit 
 partitioning of the investigated surface into $2n$ parts.\ Doing so  we 
 also increase our  chances  of  getting  smaller  overestimations.\  Of 
 course, in practice the explicit form of $f$ is also important  --  the 
 SUE's (Single Usage Expressions), if possible at all, are preferred  as 
 usually. The thin trial boxes should be constructed  with  care:  their 
 edges have to be rounded outwards.

\appendix

    It is likely that $NP$-hardness of many interval algorithms is among 
 the key factors preventing their wide  dissemination,  no  matter  that 
 they also deliver only highest quality, verified results. The algorithm 
 presented here  is  different:  its  complexity  per  single  candidate 
 scales linearly with the dimensionality of the  problem.  Hence  it  is 
 able to outperform any of its classical  counterpart  based  on  matrix 
 operations.\ Moreover, it is  simple  and  makes  no  implicit  use  of 
 unfounded  assumptions,  like  the  one  that  every  minimum  can   be 
 approximated by a quadratic form.\ In addition, it will  never  produce 
 false results. Our solution is one more example of the  old  truth:  we 
 need algorithms designed from the very beginning as interval-oriented.

\begin{acknowledgements}
    This work is part of author's statutory activities at  the  Institute 
 of Physics, Polish Academy of Sciences.
\end{acknowledgements}

\end{article}

\begin{thebibliography}{}

\bibitem[\protect\citeauthoryear{Korn\&Korn}{1968}]{Korn}
G.A.~Korn and T.M.~Korn.
\newblock {Mathematical  Handbook}.
\newblock McGraw-Hill Book Co., New York, San Francisco, Toronto,  London, Sydney (1968), ch.~13.5

\bibitem{intro}
    For a~concise and nice introduction into interval  calculations  and 
 methods visit the website \url{http://www.cs.utep.edu/interval-comp/}


\end{thebibliography}
\end{document}